\documentclass[runningheads]{llncs}
\usepackage[T1]{fontenc}

\usepackage{graphicx}
\usepackage{mathtools}
\usepackage{amsfonts}
\usepackage{newtxtext,newtxmath}

\DeclareMathAlphabet{\mathbb}{U}{msb}{m}{n}
\begin{document}
\title{Home Spaces and Invariants to Analyze Parameterized Petri Nets}
%
%
\author{Gerard Memmi
}
\authorrunning{G. Memmi}
%
\institute{LTCI, Telecom-Paris, Institut polytechnique de Paris, Palaiseau, France
}
\maketitle              
\begin{abstract}
This article focuses on comparing the notions of home spaces and invariants, in Transition Systems and more particularly, in Petri Nets as well as a variety of derived Petri Nets. 

After recalling basic notions of Petri Nets and semiflows, we then discuss important characteristics of finite generating sets for 
$\mathcal{F}$, the set of all semiflows with integer coordinates of a given Petri Net. Then, we  particularly focus on $\mathcal{F^{+}}$ the set of semiflows with non-negative coordinates. 

Minimality of semiflows and minimality of supports are critical to develop effective analysis of invariants and behavioral properties of Petri Nets such as boundedness or even liveness. 
We recall known decomposition theorems considering semirings such as 
$\mathbb{N}$ or 
$\mathbb{Q^+}$ 
then fields such as $\mathbb{Q}$.
The result over $\mathbb{N}$ is being improved into a necessary and sufficient condition.

In addition, we present general new results about the topology and the behavioral properties of a Petri Net, illustrating the importance of considering semiflows with non-negative coordinates. 


Then, we regroup a number of results around the notion of home space and home state applied to transition systems. Home spaces and semiflows are used to efficiently support the analysis of behavioral properties. 

In this regard, we present a methodology to analyze a Petri Nets by successive refinement of home spaces directly deduced from semiflows and apply it to analyze a parameterized example drawn from the telecommunication industry underlining the efficiency brought by using minimal semiflows of minimal supports as well as the new results on the topology of the model. This methodology is better articulated than in previous papers, and brings us closer to an automated analysis.
\end{abstract}

\section{Introduction}


\subsection{Motivations}
Parallel programs, distributed systems, telecommunication networks, cyber-physical systems are complex entities to design, model, and verify. Using formal verification at different stages of the system development life cycle is a strong motivation that spread throughout this paper and provide us with rationale for many concepts, definitions, and behavioral properties.

How to best define and use invariants to analyze the behavior of a model? Can we not only prove that a formula is an invariant, but find a way in which they can be organized or concisely described, a way in which they can be discovered or computed? How can invariants be combined to represent a meaningful behavior? How can invariants be decomposed into simpler and verifiable properties? How to determine whether a given decomposition is more effective than another one with regard to formal verification?

This paper which must be considered as a continuation of the work described in \cite{M23}. We want to show how linear algebra or algebraic geometry can efficiently sustain invariant calculus. 
In such a setting, linear algebra can also be applied and utilized to prove a large variety of behavioral properties. When the model is parameterized, it can be used to determine in which domain behavioral properties can be satisfied.




\subsection{Outline and contributions}
After providing some basic notations in Section \ref{sec: basic notations} and recalling a first set of classic properties for semiflows (be in $\mathbb{Z}$ or $\mathbb{N}$) in Section \ref{sec: semi}, the notions of generating sets are briefly recalled from \cite{M23}.



The notions of generating sets, minimal semiflows, and minimal supports are then defined in Section \ref{sec: generating sets}. 

The three decomposition theorems of Section \ref{sec: 3theorems} have been first published in \cite{M78} then slightly improved in \cite{M23}. 
Here, the first theorem is extended one more time to fully characterize minimal semiflows and generating sets over $\mathbb{N}$. The other two theorems are just recalled to present a complete result.
Subsequently, theorem \ref{th: bounds} describes three extremums regarding any semiflows and place in a support of a semiflow. This result can be computed from any generating set. This important detail was never stressed out before despite its importance from a computational point of view. This will later be used in the analysis of two examples presented Section \ref{sec: ex}.
 


Then, the notion of home space is described Section \ref{sec: HS} with a set of results linked to their structure and later to their key relation with liveness Section \ref{subsec: hs-semiflow-liveness}. 

Two parameterized examples are given to illustrate how invariants and home spaces can be used to prove behavioral properties of a Petri Net in Section \ref{sec: ex}. 
A first example is analyzed with basic arithmetic reasoning.
A second example is drawn from the telecommunication industry Section \ref{subsec: example-summary}.
It illustrates a method of analysis by home spaces refinement based on some of the theoretical results presented in this paper; in particular theorem \ref{th: bounds}.

Section \ref{sec: concl} concludes and provides possible avenues of future research.
At last, some additional thoughts will conclude this note.

\section{Basic notations}
\label{sec: basic notations}

In this Section, we briefly recall Petri Nets, including the notion of potential state space usual in Transition Systems, introducing notations that will be used in this paper. Then, we define semiflows in $\mathbb{Z}$ and basic properties in $\mathbb{N}$ highlighting why semiflows in $\mathbb{N}$ may be considered more useful to analyze behavioral properties.

A \textit{Petri Net} is a tuple $PN = \langle P, T, Pre, Post \rangle$, where $P$ is a finite set of \textit{places}, $T$ a finite set of \textit{transitions} such that $P \cap T = \text{\O}$. 
A transition $t$ of $T$ is defined by its $Pre(\cdot,t)$ and $Post(\cdot,t)$ \textit{conditions} 
\footnote{We use here the usual notation: $Pre(\cdot,t)(p) = Pre(p,t)$ and  $Post(\cdot,t)(p) = Post(p,t)$.}:
$Pre: P \times T \rightarrow \mathbb{N}$ is a function providing a weight for pairs ordered from places to transitions, $Post: P \times T \rightarrow \mathbb{N}$ is a function providing a weight for pairs ordered from transitions to places. 
$d$ will denote the number of places: $d = |P|$.

The dynamic behavior of Petri Nets is modeled via markings. A \textit{marking or state} (as in any transition system) $q:\ P \rightarrow \mathbb{N}$ is a function that evolves with the execution (or \textit{firing}) of transitions or sequence of transitions (i.e. a word in $T^*$). When $q(p) = k$, it is often said that the place $p$ contains $k$ \textit{tokens}. 

We also define $Q$, the set of all \textit{potential markings} (also known as state space in Transition Systems); for Petri Nets we usually have: $Q = \mathbb{N}^d$. Then, $RS(PN,A)$ will denote the reachability set of a Petri Net (or model) PN from a subset $A$ of $Q$: $RS(PN,A) = \{q \in Q \ |\  \exists a \in A,\ q\  \texttt{reachable from}\ a\}$. $RG(PN,A)$ or $LRG(PN,A)$ will denote the corresponding reachability graph with labels in $T$ for $LRG(PN,A)$, without labels in $T$ for $RG(PN,A)$ as in Figure \ref{fig: inter-hs}.

Extensive definitions, properties, and case studies can be found in many lecture notes and books, in particular in \cite{BR82,GV03}.

\section{Petri Nets and Semiflow basic properties}
\label{sec: semi}

\begin{definition}
A \textit{semiflow} is a solution of the following homogeneous system of $|T|$ diophantine equations: 
\begin{equation}
\label{eq:inv-semiflow}
    f^\top Post(\cdot,t) = f^\top Pre(\cdot,t), \ \ \forall t \in T,
\end{equation}
where $x^\top y$ denotes the scalar product of the two vectors $x$ and $y$ since $f, Pre$ and $Post$ can be considered as vectors once the places of $P$ have been ordered. 
\newline
$\mathcal{F}$ and $\mathcal{F}^+$ denote the sets of solutions of the system of equations (\ref{eq:inv-semiflow}) which have their coefficients in $\mathbb{Z}$ and in $\mathbb{N}$ respectively.
\end{definition}

Considering a Petri Net $PN$ with its initial marking $q_0$ and the set of reachable markings from $q_0$ through all sequences of transitions denoted by $RS(PN,q_0)$, any non-null solution $f$ of the homogeneous system of equations (\ref{eq:inv-semiflow}) allows to directly deduce the following \textit{invariant} of the Petri Net defined by its $Pre$ and $Post$ functions (used in the system of equations that $f$ satisfies):
\begin{equation}
\label{eq:inv}
\forall \ q\ \in RS(PN,q_0): 
    f^\top q = f^\top q_0.
\end{equation}
In the rest of the paper, we abusively use the same symbol ‘0' to denote 
$(0,...,0)^\top$ of $\mathbb{N}{^n},\ \forall n \in \mathbb{N}$. 
The \textit{support} of a semiflow $f$ is denoted by $\left \| f \right \|$ and is defined by:

$\left \| f \right \| = \{x\in P\ |\ f(x) \neq 0\}$.

We will use the usual componentwise 
partial order in which

$(x_1, x_2, \dots, x_d )^\top \leq ( y_1, y_2, \dots, y_d )^\top$ if and only if $x_i \leq y_i\ \forall i \in \{1, \dots, d\}$.

However, the most interesting set of semiflows from a behavioral analysis point of view are defined over natural numbers instead of and is denoted by $\mathcal{F}^+$. This can be seen through the three following properties. 
First, we define the \textit{positive and negative supports} of a semiflow $f \in \mathcal{F}$ as:

$\left\| f\right\|_+ = \left\{ p\in P\ |\ f(p)> 0\right\}$,

$\left\| f\right\|_- = \left\{ p\in P\ |\ f(p)< 0\right\}$,

with $\left\| f\right\| = \left\| f\right\|_- \cup\   \left\| f\right\|_+$.

We can then rewrite the system of equations (\ref{eq:inv}):

\begin{equation}
\label{eq: differential}
f^\top q=\left| \sum_{p \in \left\| f\right\|_+}f(p)q(p)\right| - \left| \sum_{p \in \left\| f\right\|_-}f(p)q(p)\right|=f^\top q_0
\end{equation}

As we can see, the formulation of equation \ref{eq: differential} is a differential between the weighted number of tokens in the places of the positive support and the weighted number of tokens in the places of the negative support of $f$.
This differential allows deducing an invariant since by equations (\ref{eq: differential}) it remains constant during the evolution of the Petri Net. 
A first general property can be immediately deduced recalling that the initial state $q_0$ belongs to $\mathbb{N}^d$.

\begin{property}
\label{prop: boundedness in F}
    For any semiflow $f \in \mathcal{F}$,
    $\exists p \in \left\| f\right\|_+$ not bounded if and only if $\exists p \in \left\| f\right\|_-$ not bounded.
\end{property}

Of course, if $\left\| f\right\|_- = \varnothing$ then $f \in \mathcal{F}^+$ and $\left\| f\right\|$ is necessarily structurally bounded. 
More generally, considering a weighting function $f$ over $P$  being defined over non-negative integers, the following properties can be easily proven \cite{M78}:  
\begin{property}
\label{prop: boundedness-alg}
If $f \geq 0$ is such that $f^{T} Pre(\cdot ,t) \geq  f^{T} Post(\cdot ,t)\ \forall t \in T$,
then the set of places of $\left\|f \right\|$ is structurally bounded (i.e. bounded from any initial marking). 

Moreover, the marking of any place $p$ of $\left \| f \right \|$ has an upper bound:

$q(p) \leq \frac{f^{T}q_0}{f(p)}, \  \ \forall q \in RS(PN,q_0)$.
\end{property}

If $f > 0$ then $\left\|f \right\|_+ = \left\|f \right\| = P$ and the Petri Net is also structurally bounded. The reverse is also true: if the Petri Net is structurally bounded, then there exists a strictly positive solution for the system of inequalities above (see \cite{Si78} or \cite{BR82}). 
This property is indeed false for a semiflow that verifies the system but would have at least one negative element and constitutes a first reason for particularly considering weight functions $f$ over $P$  being defined over non-negative integers including $\mathcal{F}^+$. 

The following corollary can directly be deduced from the fact that any semiflow in $\mathcal{F}^+$ satisfies property \ref{prop: boundedness-alg}:

\begin{corollary}
\label{prop: bound-for-support}
    For any place $p$ belonging to at least one support of a semiflow of $\mathcal{F}^+$, an upper bound $\mu$ can be defined for the marking of $p$ such that:   

    $\forall q \in RS(PN,q_0), \ \ q(p) \leq \mu(p,q_0) =  \min_{\{f\in \mathcal{F}^+\ |\ f(p) \neq 0\}} \frac{f^\top q_0}{f(p)} \ $
\end{corollary}
A second reason for particularly considering a semiflow $f$
as being defined over non-negative integers is that the system of inequalities: 

\begin{equation}
 f^{T} q_0  \geq f^{T} Pre(\cdot,t), \ \ \ \ 
 \forall t \in T
\label{eq:threshold}
\end{equation}
  
becomes a necessary condition for any transition $t$ to stand a chance to be enabled from any reachable marking from $q_0$, then to be live. In \cite{BR82}, $f^{T} Pre(\cdot,t)$ is called the \textit{enabling threshold} of $t$. 
\begin{property}
\label{prop: enabling-threshold}
If $t$ is a transition and $\exists f \in \mathcal{F}^+ \setminus \{0\}$ such that:

 $f^{T} q_0  < f^{T} Pre(\cdot,t)$, then $t$ cannot be executed from $\langle M,q_0 \rangle$.
\end{property}
This property can be used when looking for a frugal management of resources (i.e. a marking as small as possible) and still fully functioning (i.e. live). Another application is when the model is defined with parameters, then, some values of these parameters for which the model is not live (see example of figure \ref{tinyk}) can be directly deduced from theorem \ref{th: bounds}.

At last, the following property can easily be proven true in $\mathcal{F}^+$ and not true in $\mathcal{F}$: 
\begin{property}
\label{prop: support-plus-union}
If $f$ and $g$ are two semiflows with no negative coefficients, then we have:
$\left\|f+g \right\|=\left\|f \right\|\cup \left\| g\right\|$.

If $\alpha$ is a non null integer then $\left\|\alpha f \right\|=\left\|f \right\|$.

\end{property}
This  property can already be found in \cite{M78} or \cite{BR82} . 

These results have been cited and utilized many times in various applications going beyond computer science, electrical engineering, or software engineering. 
For instance, they have been used in the domain of biomolecular chemistry relatively to chemical reaction networks \cite{JACB18} which brings us back to the original vision of C. A. Petri when he highlighted that his nets could be used in chemistry. Many other applications can be found in the literature. 

\section{Generating sets and minimality}
\label{sec: generating sets}

\subsection{Generating sets}

The notion of generating sets for semiflows is well known and efficiently supports the handling of an important class of invariants. 
Several results have been published starting from the initial definition and structure of semiflows  \cite{M77} to a large array of applications used especially to analyze Petri Nets \cite{Colom2003,DworLo16,JACB18,Wol2019}.

Minimality of semiflows and minimality of their supports are critical to understand how to best decompose semiflows and manage analysis of behavioral properties.
Invariants directly deduced from minimal semiflows relate to smaller quantities of resources.
Furthermore, the smaller the support of semiflows, the more local their footprint (i.e. the smaller the number of resources).
In the end, these two notions of minimality will foster analysis optimization.

\subsection{Basic definitions and results}
\label{subsec: gs-basic-results}

\begin{definition}
\label{def: FG}
A subset $\mathcal{G}$ of $\mathcal{F^{+}}$ is a \textit{generating set over a set $\mathbb{S}$} (where $\mathbb{S} \in \{\mathbb{N}, \mathbb{Q^{+}}, \mathbb{Q}\}$ with $\mathbb{Q^{+}}$ denoting the set of non-negative rational numbers) if and only if $\forall f \in \mathcal{F}^+$ 
we have $f = \sum_{g_i \in \mathcal{G}} \alpha_ig_i $ where $\alpha_i \in \mathbb{S}$ and $g_i \in \mathcal{G}$.
\end{definition}

Since $\mathbb{N} \subset \mathbb{Q^{+}} \subset \mathbb{Q}$, \footnote{ Where $\subset$ denotes the strict inclusion between sets.} a generating set over $\mathbb{N}$ is also a generating set over $\mathbb{Q^{+}}$, and a generating set over $\mathbb{Q^{+}}$ is also a generating set over $\mathbb{Q}$. 
However, the reverse is not true and is, in our opinion, a source of some inaccuracies that can be found in the literature. 
Therefore, it is important to precise over which set the coordinates (used for the decomposition of a semiflow) belong.






\textbf{-}If $\mathcal{G}$ is a generating set over $\mathbb{Q}$ then 

$\mathcal{V(G)} = \{ f \in \mathbb{Q}{^d}\ |\ f = \sum_{i=1}^{i=k} \alpha_ig_i\ \text{and}\ \alpha_i \in \mathbb{Q}\}$ is a \textit{vector space} and $\mathcal{F^{+}} = \mathcal{V(G)} \cap \mathbb{N}{^d}$. We can extract from $\mathcal{G}$ a basis of $\mathcal{V(G)}$ (see, for instance, \cite{Lan02} p. 85) which also is a generating set of $\mathcal{F^{+}}$ over $\mathbb{Q}$ since the elements of this basis are in $\mathcal{F^{+}}$.

\subsection{Minimal supports and minimal semiflows}
The fact that there exists a finite generating set over $\mathbb{N}$ is non trivial. This result was proven by Gordan circa 1885 then Dickson circa 1913. Here, we directly rewrite Gordan's lemma \cite{AB86} by adapting it to our notations.

\begin{lemma} (\textbf{Gordan circa 1885})
\label{lem: Gordan}
Let  $\mathcal{F^+}$ be the set of non-negative integer solutions of the system of equations (\ref{eq:inv-semiflow}). Then, there exists a finite generating set of vectors in $\mathcal{F^+}$ such that every element of $\mathcal{F^+}$ is a linear combination of these vectors with non-negative integer coefficients.
\end{lemma}

The question of the existence of a finite generating set being solved for $\mathbb{N}$, it is necessarily solved for $\mathbb{Q^+}$ and $\mathbb{Q}$. This lemma is necessary not only to prove the decomposition theorem but also to claim the computability of the extremums described in Theorem \ref{th: bounds}.

Several definitions around the concept of minimal semiflow were introduced in \cite{STC1998} p. 319, in \cite{ColomTS2003} p. 68, \cite{Kruck86}, \cite{CMPW09}, or in \cite{M78,M83}, however, we will only consider
two basic notions in order theory: minimality of support with respect to set inclusion and minimality of semiflow with respect to the componentwise partial order on $\mathbb{N}{^d}$ since the various definitions we found in the literature as well as the results of this note can be described in terms of these sole two classic notions.

\begin{definition}[minimal support]
A non-empty support $ \left \| f \right \| $ 
of a semiflow $f$ is \textit{minimal} with respect to set inclusion if and only if $\nexists \ g \in \mathcal{F}^+\setminus\{0\}$ such that $\left \| g \right \| \subset \left \| f \right \| $. 
\end{definition}

\begin{definition}[minimal semiflow]
A non-null semiflow $f$ is \textit{minimal} with respect to $\leq$ if and only if $\nexists \ g \in \mathcal{F^{+}}\setminus\{0,f\}$ such that $g \leq f$.
\end{definition}

In other words, a minimal semiflow cannot be decomposed  as the sum of another semiflow and a non-null non-negative vector. 
This remark yields initial insight into the foundational role of minimality regarding the decomposition of semiflows.
We are looking for characterizing generating sets such that they allow analyzing various behavioral properties as efficiently as possible.

The notion of minimality is key to decompose semiflows then to analyze behavioral properties concisely. First, if we were to consider a generating set over $\mathbb{N}$, then we may have to explore every minimal semiflow. Although finite, the number of minimal semiflows can be quite large. Second, considering a basis over $\mathbb{Q}$ is of course relevant to handle $\mathcal{F}$ less when it is about $\mathcal{F}^+$ and may not capture behavioral constraints as easily (see the example of Section \ref{subsec: example-summary}).

\subsection{Three decomposition theorems}
\label{sec: 3theorems}

Generating sets can be characterized thanks to a set of three decomposition theorems. A first version of them can be found in \cite{M78} with their proofs. A second version version can be found in \cite{M23} with few improvements.  
Here, only theorem \ref{th: over N} which is valid over $\mathbb{N}$   
is extended to fully characterize minimal semiflows and generating sets over $\mathbb{N}$, and is provided with a new proof using Gordan's lemma \ref{lem: Gordan}.
Theorems \ref{th: min support} and \ref{th: decomp} are recalled for completeness and are unchanged from \cite{M23}.

\subsubsection{Decomposition over non-negative integers}

\begin{theorem} (\textbf{Decomposition over $\mathbb{N}$})
\label{th: over N}

A semiflow is minimal if and only if it belongs to any generating set over $\mathbb{N}$.

The set of minimal semiflows of $\mathcal{F^{+}}$ is a finite generating set over $\mathbb{N}$.
\end{theorem}
Let's consider a semiflow $f \in \mathcal{F^{+}}\setminus\{0\}$ and its decomposition over any family of $k$ non-null semiflows $f_i, 1 \leq i \leq k$. Then, $\exists a_1,...,a_k \in \mathbb{N}$ 
such that $f = \sum_{i=1}^{i=k}a_if_i$. 
Since $f \neq 0$ and all coefficients $a_i$ are in $\mathbb{N}$, $\exists j \leq k$ such that $0< f_j \leq a_jf_j \leq f$. If $f$ is minimal, then $a_j=1$ and $f_j=f$.
Hence, if a semiflow is minimal, then it belongs to any generating set over $\mathbb{N}$. The reverse will become clear once the second statement of the theorem is proven.

Applying Gordan's lemma, there exists a finite generating set, $\mathcal{G}$ \footnote{This point is taken for granted in \cite{M78} as well as the rest of literature on semiflows.}. Since any minimal semiflow is in $\mathcal{G}$, the subset of all minimal semiflows is included in $\mathcal{G}$ and therefore finite. Let $\mathcal{E} = \{e_1,...e_n\}$ be this subset and prove by construction that $\mathcal{E}$ is a generating set. 

For any semiflow $f \in \mathcal{F^+}$, we 
build the following sequence leading to the decomposition of $f$:

i) $r_0 = f$

ii)  $r_i = r_{i-1} - k_ie_i$ such that $ r_i \in  \mathcal{F^+}$
and $r_{i-1} - (k_i+1)e_i \notin \mathcal{F^+}$

By construction of the non-negative integers $k_i$ , we have $r_n \in \mathcal{F^+}$ and $\nexists e_i \in \mathcal{E}$ such that $e_i \leq r_n$. 
This means that $r$ is either minimal or null. Since $\mathcal{E}$ includes all minimal semiflows, therefore, 
$r=0$, and any semiflow can be decomposed as a linear combinations of minimal semiflows, in other words, $\mathcal{E}$ is a finite generating set.
\footnote{If $\mathcal{E}$ was to be  infinite, the construction could still be used since the monotonically decreasing sequence $r_i$ is bounded by 0 and $\mathbb{N}$ is nowhere dense, so we would have:
\newline
$ \lim \limits_{n \to \infty} f- \sum_{j=1}^{j=n}k_je_j = 0$ with the same definition of the coefficients $k_j$ as in ii).}
It is now clear that if a semiflow $f$ belongs to any generating set, then it belongs in particular to $\mathcal{E}$, therefore, $f$ is a minimal semiflow. 
\hfill
$\square$

Let's point out that since $\mathcal{E}$ is not necessarily a basis, the decomposition is not unique in general and depends on the order in which the minimal semiflows of $\mathcal{E}$ are considered to perform the decomposition.

However, a minimal semiflow does not necessarily belong to a generating set over $\mathbb{Q^{+}}$ or $\mathbb{Q}$. 

\subsubsection{Decomposition over semiflows of minimal support}
This theorem \ref{th: min support} can already be found in \cite{M23}.
\begin{theorem} (\textbf{Minimal support})
\label{th: min support}
If $I$ is a minimal support then 

i) there exists a unique minimal semiflow $f$ such that $I = \left \| f \right \|$ and $\forall g \in \mathcal{F^{+}}$ such that $\left \| g \right \| = I, \exists k \in \mathbb{N}$ such that $g = kf$,

ii) any non-null semiflow $g$ such that $\left \| g \right \| = I$ constitutes a generating set over $\mathbb{Q^{+}}$ or $\mathbb{Q}$ for $\mathcal{F}_I^+ = \left\{ g \in \mathcal{F^+} |\ \left\| g\right\|= I\right\}$.
\end{theorem}
In other words, $\{f\}$ is a unique generating set over $\mathbb{N}$ for $\mathcal{F}_I^+ =\{g \in \mathcal{F^{+}} \ | \ \left \| g \right \| = I\}$. 
However, this uniqueness property is indeed lost in $\mathbb{Q^{+}}$ or in $\mathbb{Q}$, since any element of $\mathcal{F}_I^+$ is a generating set of $\mathcal{F}_I^+$ over $\mathbb{Q^{+}}$ or $\mathbb{Q}$.

\begin{theorem} (\textbf{Decomposition over $\mathbb{Q}^+$})
\label{th: decomp}
Any support $I$ of semiflows is covered by the finite subset $\{I_1, I_2, \dots, I_N\}$ of minimal supports of semiflows included in $I$:

$I = \bigcup_{i=1}^{i=N} I_i$.

Moreover,
$\forall f \in \mathcal{F^{+}}$ such that $\left \| f \right \| \subseteq  I$, one has $f=\sum_{i=1}^{i=N} \alpha_ig_i$ where $ \forall i \in \{1,2,...N\},\ \alpha_i \in \mathbb{Q^{+}}$ and the semiflows $g_i$ are such that $\left \| g_i \right \| = I_i$.
\end{theorem}

A sketch of proof of theorem \ref{th: decomp} can be found in \cite{BR82}, a complete proof in \cite{M78}.

\subsection{Three extremums drawn from the notion of semiflow}
\label{subsec: bounds}
The knowledge of any finite generating set allows a practical computation of the three extremums directly inspired from property \ref{prop: inter-home-spaces} Section \ref{subsec: hs-semiflow-liveness}, corollary \ref{prop: bound-for-support} and property \ref{prop: support-plus-union} of Section \ref{sec: semi}. Let us define them:
\begin{definition}
Given an initial state $q_0$ and the set of semiflows $\mathcal{F^+}$, the three following limits can be defined:
\begin{itemize}
\item $\iota = \bigcap_{f \in \mathcal{F^+}} HS(f,q_0)$, where $HS(f,q_0) = \{q \in Q\ |\ f^\top q = f^\top q_0 \}$,
\item  
$\mu(p,q_0) = \min_{\{f\in \mathcal{F}^+\ |\ f(p) \neq 0\}} \frac{f^\top q_0 }{f(p)}$ is the lowest bound that can be built from a semiflow the support of which contains the given place $p$ in $P$,
\item $\rho = \left\| \sum_{f\in \mathcal{F}^+} f\right\|$ is the largest support of any semiflow in $\mathcal{F^+}$.
\end{itemize}
\end{definition}

Theorem \ref{th: bounds} expresses the fact that these extremums are computable as soon as any generating set is available: 
\begin{theorem}
\label{th: bounds}
Let's $\mathcal{E} = \{e_1,...e_N\}$ be any finite generating set of $\mathcal{F}^+$, and $q_0 \in Q$ an initial state:
\begin{itemize}
  \item If $\mathcal{E}$ is over $\mathbb{S}$ then we have:
  
  $\iota = \bigcap_{f \in \mathcal{F^+}} HS(f,q_0) = \bigcap_{e_i \in \mathcal{E}} HS(e_i,q_0)$,
  \item If $\mathcal{E}$ is over $\mathbb{Q}^+$ or $\mathbb{N}$ then for any place $p$ belonging to at least one support of a semiflow of $\mathcal{F}^+$, $\forall q \in RS(PN,q_0)$, we have :   

$q(p) \leq \mu(p,q_0) = \min_{\{f\in \mathcal{F}^+\ |\ f(p) \neq 0\}} \frac{f^\top q_0}{f(p)} = \min_{\{e_i\in \mathcal{E}\ |\ e_i(p) \neq 0\}} \frac{{e_i}^\top q_0}{e_i(p)} $,

  \item If $\mathcal{E}$ is over $\mathbb{S}$ then we have:

$\rho = \left\| \sum_{f\in \mathcal{F}^+} f\right\|= \bigcup_{f\in \mathcal{F}^+}\left\|f\right\| = \bigcup_{e_i \in \mathcal{E}}\left\|e_i\right\|$
 
\end{itemize}
\end{theorem}

First, the three extremums are computable since by Gordan's lemma of section \ref{subsec: gs-basic-results}, there exists a finite generating set such as $\mathcal{E}$.
\begin{itemize}
    \item 

For the first item, let's consider: 

$f \in \mathcal{F^+}$ with $f = \sum_{i=1}^{i=N} \alpha_ie_i$ and $q \in \bigcap_{e_i \in \mathcal{E}} HS(e_i,q_0)$, then:

$\alpha_i(e_i^\top q) = \alpha_i(e_i^\top q_0) \ \forall i\in \{1,...N\}$, hence:  

$ \sum_{i = 1}^{i = N} \alpha_i(e_i^\top q) = \sum_{i = 1}^{i = N} \alpha_i(e_i^\top q_0)$, then:

$f^\top q = f^\top q_0$, 
and $q \in HS(f,q_0)\ \ \forall f \in \mathcal{F^+}$ therefore, 

(since $\mathcal{E} \subset \mathcal{F^+}$ directly involves $(\bigcap_{f \in \mathcal{F^+}} HS(f,q_0))  \subseteq \bigcap_{e_i \in \mathcal{E}} HS(e_i,q_0)$) we have: 

$\bigcap_{e_i \in \mathcal{E}} HS(e_i,q_0) = \bigcap_{f \in \mathcal{F^+}} HS(f,q_0) = \iota$.
\item
For the second item of the theorem, let's consider a state $q_0$, a place $p$, and a semiflow $f$ of $\mathcal{F}^+$ such that $f(p) > 0$ and $f = \sum_{i=1}^{i=N}\alpha_i e_i$ where $\alpha_i \geq 0\ \forall i \in \{1,...N\}$.

Let's 
define $\mu_\mathcal{E}$ such that:
$\mu_\mathcal{E} = 
\min_{\{e_i\in \mathcal{E}\ |\ e_i(p) \neq 0\}} \frac{{e_i}^\top q_0}{e_i(p)}$.

Then $\exists j$ such that $1 \leq j \leq N$, and $\mu_\mathcal{E} = \frac{{e_j}^\top q_0}{e_j(p)}$.

Therefore, $\forall i \leq N$, such that $e_i(p) \neq 0$,  
$\exists \delta_i \in \mathbb{Q}^+$ such that:
$\frac{{e_j}^\top q_0}{e_j(p)} = \frac{{e_i}^\top q_0 - \delta_i}{e_i(p)}$. It can then be deduced:

$\mu_\mathcal{E} = \frac{\alpha_j{e_j}^\top q_0}{\alpha_j e_j(p)} = \frac{\alpha_i({e_i}^\top q_0 - \delta_i)}{\alpha_i e_i(p)}\ \forall i$ such that $e_i(p) \neq 0$, therefore:

$\mu_\mathcal{E} = \frac{\sum_{\{i\ |\ e_i(p) > 0\}} \alpha_i({e_i}^\top q_0 - \delta_i)}{\sum_{\{i\ |\ e_i(p) > 0\}}\alpha_i e_i(p)}$

$= \frac{\sum_{\{i\ |\ e_i(p) > 0\}} \alpha_i({e_i}^\top q_0 - \delta_i) + \sum_{\{i\ | e_i(p) = 0\}}(\alpha_i{e_i}^\top q_0 - \alpha_i{e_i}^\top q_0)} {\sum_{\{i\ |\ e_i(p) > 0\}}\alpha_i e_i(p)} $

$= \frac{\sum_{\{i | e_i(p) > 0\}} \alpha_i{e_i}^\top q_0 + \sum_{\{i | e_i(p) = 0\}}\alpha_i{e_i}^\top q_0 - \sum_{\{i | e_i(p)>0\}}\alpha_i\delta_i - \sum_{\{i\ |\  e_i(p) = 0\}}\alpha_i{e_i}^\top q_0} {\sum_{\{i\ |\ e_i(p) > 0\}}\alpha_i e_i(p) + \sum_{\{i\ |\ e_i(p) = 0\}}\alpha_i e_i(p)}$

since $\sum_{\{i\ |\ e_i(p) = 0\}}\alpha_i e_i(p) = 0$. Then, since $\delta_i \geq 0$ and $\alpha_i \geq 0\  \forall i$ such that $1\leq i \leq N$

$\mu_\mathcal{E} = \frac{\sum_{i=1}^{i=N}\alpha_i{e_i}^\top q_0 -\sum_{\{i | e_i(p)>0\}}\alpha_i\delta_i - \sum_{\{i\ |\  e_i(p) = 0\}}\alpha_i{e_i}^\top q_0}{\sum_{i=1}^{i=N}\alpha_i e_i(p)}$

$\mu_\mathcal{E} = \frac{f^\top q_0 -\sum_{\{i | e_i(p)>0\}}\alpha_i\delta_i - \sum_{\{i\ |\  e_i(p) = 0\}}\alpha_i{e_i}^\top q_0}{f(p)} \leq  \frac{f^\top q_0 }{f(p)}$ 

This being verified for any semiflow of $\mathcal{F}^+$, we have: $\mu(p,q_0) = \mu_\mathcal{E}$.

\item 
For the third item of the theorem, let's consider $\mathcal{E}$ a generating set over $\mathbb{S}$ then, any semiflow $f$ can be decomposed as follows:

$f = \sum_{\alpha_i > 0} \alpha_ie_i + \sum_{\alpha_i < 0}\alpha_ie_i$ where $\alpha_i \in \mathbb{S}$.
$f \in \mathcal{F}^+$ means that at least one coefficient $\alpha_i$ is strictly positive and $\sum_{\alpha_i < 0}|\alpha_i|e_i + f = \sum_{\alpha_i > 0}\alpha_ie_i \neq 0$.

Therefore, applying property \ref{prop: support-plus-union}: 

$\left\|f\right\| \subseteq \left\|\sum_{\alpha_i < 0}|\alpha_i|e_i + f\right\| = \left\|\sum_{\alpha_i > 0} \alpha_ie_i\right\| = \bigcup_{\alpha_i > 0} \left\|\alpha_ie_i\right\| \subseteq \bigcup_{e_i \in \mathcal{E}}\left\|e_i\right\|$.

Hence, $\rho = \left\| \sum_{f\in \mathcal{F}^+} f\right\|= \bigcup_{e_i \in \mathcal{E}}\left\|e_i\right\|$
\hfill
$\square$
\end{itemize}

This theorem means that these three extremums $\iota$, $\mu$ and $\rho$ can be computed with the help of one finite generating set. 
The third part of this theorem means that if $\mathcal{E} = \{e_1,...e_N\}$ is any generating set of a given Petri Net PN then $\bigcup_{e_i \in \mathcal{E}}\left\|e_i\right\|$ is also the largest support of PN (it would be useless to look for another semiflow).

\section{Home spaces and home states}
\label{sec: HS}

The notion of home space was first defined in \cite{M83} for Petri Nets relatively to a single initial state. Here, we effortlessly extend its definition relatively to a nonempty subset of states. 
Verifying that a set of states is a home space is possible as soon as a conceptual model supports building up Reachability Sets ($RS$) and associated reachability graphs ($RG$ and $LRG$ for Reachability Graph, and Labeled Reachability Graph respectively). 

Home spaces are extremely useful to analyze liveness, or resilience (see \cite{FinHil24}). Any behavioral property requiring to eventually become satisfied after executing a known sequence of transitions can be supported by a home space (a property satisfied for any marking or state would be an invariant). 

\subsection{Definitions and basic properties}
\label{subsec: HS-def}
Given a model $M$, its associated state space $Q$ and a subset $Init$ of $Q$, we say that a set \emph{HS} is an \textit {Init-home space} if and only if for any progression of the model $M$ from any element of $Init$, there exists a way of  prolonging this progression and reach an element of \emph{HS}. In other words:

\begin{definition}
\label{def: homespace}
Given a nonempty subset $Init$ of $Q$, a set $\emph{HS}$ is an \textit{Init-home space} if and only if $\forall q\in RS(M,Init)$, $\exists h \in \emph{HS}$ such that $h$ is reachable from $q$ (i.e. such that $q \overset{*}{\rightarrow}h$).

$\emph{HS}$ is a \textit{well-structured home space} if and only if: 
$\forall q\in Q$, $\exists  h \in \emph{HS}$ such that $q \overset{*}{\rightarrow}h$.
\end{definition} 
This definition is general and can be applied to any Transition System. In \cite{JALE22}, we can find for Petri Nets, an equivalent definition : $\emph{HS}$ is an \textit{Init-home space} if and only if $RS(M,Init) \subseteq RS^{-1}(M,HS \cap Q)$.

We immediately have $HS \cap RS(M,Init) \neq \varnothing$. However, depending on the definition of $Q$ and for algebraic reasons, we must point out that $HS$ is not necessarily included in $Q$ and we may have $h \in HS$ such that for some state variables $x_i, \ h(x_i) \notin D_i$ ($D_i$, being the domain in which $x_i$ varies in $Q$).
 
If $HS$ is a well-structured home space then 
$\forall Init \in 2^Q \setminus 
\{\emptyset\}$, 
$HS$ is an \textit{Init}-home space. 
Also, if $A$ is such that $Q=RS(M,A)$ and $\emph{HS}$ is an A-home space then $\emph{HS}$ is a well-structured home space.

\begin{definition}
\label{def: home state}
A state $s$ is an \textit{home state} if and only if 
$\{s\}$ is an $\{s\}$-home space. 
\end{definition}
This last definition is directly drawn from the definition given in \cite{BR82} p.59 or in \cite{GV03} p. 63 for Petri Nets.
It can be found in many papers such as \cite{HDMK14}.

In many systems, the initial state $q_0$ represents an \textit{idle} state from which the various capabilities of the system can be enabled. In this case, it is important for $q_0$ to be a home state.
This property is usually guarantied by a reset function which can be modeled in a simplistic way by adding a transition $r$ such that $\forall q \in RS(M,q_0),\ q_0 \in r(q)$ (which means that $r$ is executable from any reachable state and that its execution reaches $q_0$). 
However, requiring to add too much complexity to $RG$ (one edge per node), this function is most of the time abstracted away when building RG up.

\begin{figure}[ht]
\centering
\includegraphics[width=0.4\textwidth]{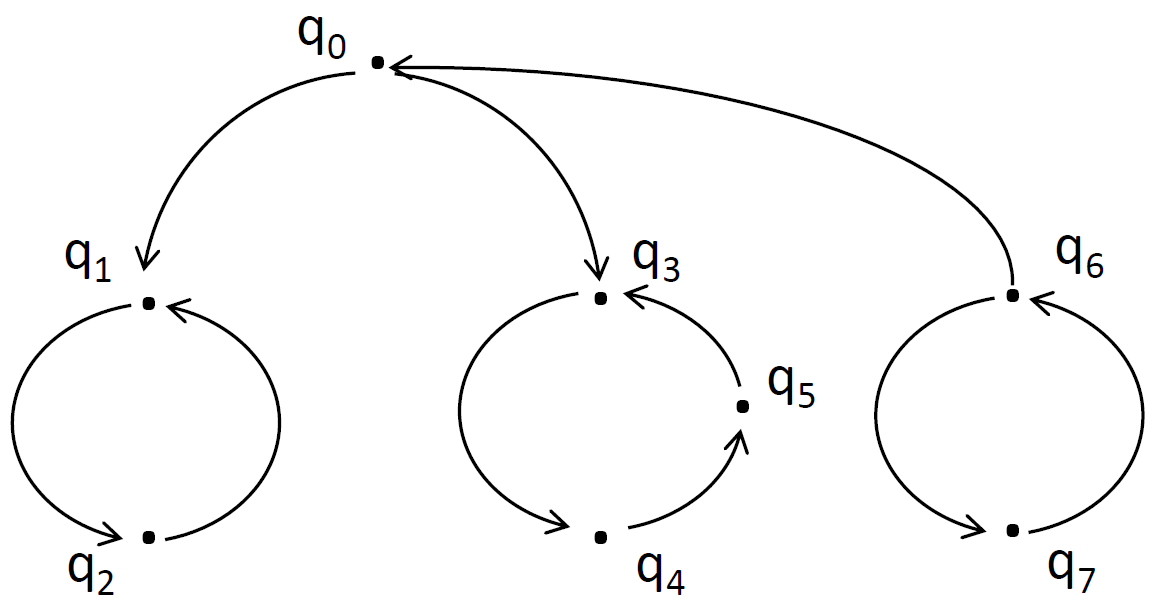}
\caption{$Q=\{q_0,q_1,q_2,q_3,q_4,q_5,q_6,q_7\},\ HS_1 = \{q_1,q_3,q_4\},\ HS_2= \{q_1, q_5\}, \ HS_3=\{q_1,q_3,q_5\}$ are three $\{q_0\}$-home space. 
$HS_4=\{q_1, q_4,q_7\}$ is a $\{q_6\}$-home space as well as a $\{q_0\}$-home space.
}
\label{fig: inter-hs}
\end{figure}

It is not always easy to prove that a given set is an A-home space. This question is addressed in \cite{JALE22} and is proven decidable for home state when the conceptual model is a Petri Net but is still open in a more general case. Furthermore, a corpus of decidability properties can be found in \cite{ValkJ85,EJ89,FinHil24}, or \cite{JALE22}. 
It may be worth mentioning the straightforward following properties, given a subset of states, $A$: 
\begin{property}
\label{prop: inter-home}
    Any set containing an A-home space is also an A-home space. If $HS$ is an A-home space, it is a B-home space for any nonempty subset $B$ of $A$. If $HS_1$ is an A1-home space and $HS_2$ is an A2-home space then $HS_1 \cup HS_2$ is an ($A1 \cup A2$)-home space.  
\end{property} 
However, the intersection of two home spaces is not necessarily a home space. Figure \ref{fig: inter-hs} is representing the reachability graph of a transition system with eight states. $HS_1, HS_2, HS_3$ as defined Figure \ref{fig: inter-hs} are three $\{q_0\}$-home spaces. 
While $HS_1 \cap HS_3 = \{q_1, q_3\}$ is a $\{q_0\}$-home space, $HS_1 \cap HS_2 = \{q_1\}$ is not a $\{q_0\}$-home space (even if it is a $\{q_1\}$-home state).
    
Given a model $M$ and a subset of states $Init$, a \textit{sink} is a state with no successor in the associated reachability graph $RG(M,Init)$.
\begin{property}
If there exists a sink $q_f \in RG(M,Init)$ then it belongs to any Init-home space.
More generally, any home space has at least one element in each strongly connected component sink of the reachability graph \footnote{It is easy to prove that this property holds even when the reachability graph is infinite considering that the definitions of sources, sinks, or strongly connected components are the same as in the case where the directed reachability graph is finite.}. 
\end{property}

For the following property, we consider a model $M$ paired with a single initial state $q_0$. 
\begin{property}
\label{home-state}
    The three following statements are equivalent: 
    \begin{itemize}
   \setlength\itemsep{0em}
    \item [(i)] the initial state is a home state (then there is no sink), 
    \item [(ii)] every reachable state is a home state, 
    \item [(iii)] the reachability graph is strongly connected.
    \end{itemize}
\end{property} 

\subsection{Home spaces, semiflows, and liveness}
\label{subsec: hs-semiflow-liveness}
Semiflows are intimately associated with home spaces and invariants and can greatly simplify the proof of fundamental properties of Petri Nets (even including parameters as in \cite{BEISW20}) such as safeness, boundedness, or more complex behavioral properties such as liveness.  
Let us provide three properties supporting this idea. 

$\texttt{Dom(t)}$ will denote the subset of states from which the transition $t$ is executable and $\texttt{Im(t)}$ the subset of states which can be reached by the execution of $t$.
\begin{property}
A transition $t$ is live if and only if \texttt{Dom(t)} is a home space.

Moreover, if \texttt{Dom(t)} is a home space then \texttt{Im(t)} is also a home space.

\end{property} 
This relationship is further supported when considering the following property regarding home states:

\begin{property}
\label{prop: home-state-liveness}
Let $M$ be a model and $q_0$ be a home state then:

any transition that is enabled from $q_0$ is live,

more generally,

a transition is live if and only if it appears as a label in $LRG(M,q_0)$.
\end{property} 

This can easily be proven directly from the definition of liveness and the property \ref{home-state} about home states .
\hfill 
$\square$

Given an initial state $q_0$, each semiflows can be associated with an invariant which in turn can be associated with a home space. In other words, if $f \in \mathcal{F}$, then $HS(f,q_0) = \{q \in Q \ |\ f^\top q = f^\top q_0\}$ is a $\{q_0\}$-home space since $RS(M,q_0) \subseteq HS(f,q_0)$. 

\begin{property}
\label{prop: inter-home-spaces}
    If $f \in \mathcal{F}$ then:
    $\forall \alpha \in \mathbb{Q} \setminus \{0\}, HS(\alpha f,q_0)= HS(f,q_0)$,

    $\forall f,g \in \mathcal{F}, \forall \alpha, \beta \in \mathbb{Q}, HS(f,q_0) \cap HS(g, q_0) \subseteq HS(\alpha f+ \beta g,q_0)$. Moreover, $HS(f,q_0) \cap HS(g, q_0)$ is a $\{q_0\}$-home space.
\end{property}
$HS(f,q_0)\ \cap \ HS(g, q_0)$ is straightforwardly a $\{q_0\}$-home space since they both contain $RS(M,q_0)$. 
Let us recall that in general, the intersection of home spaces is not a home space (see Figure \ref{fig: inter-hs}). 
If $q \in HS(f,q_0) \cap HS(g,q_0)$, then $\alpha (f^\top q) = \alpha (f^\top q_0)$ and $\beta (g^\top q) = \beta (g^\top q_0)$, 

so $(\alpha f+\beta g)^\top q = (\alpha f+ \beta g)^\top q_0$, therefore, $q \in HS(\alpha f+\beta g,q_0)$
\hfill 
$\square$

These three properties provide us with a methodology to analyze and prove that a subset of transitions are live.
From a set of invariants, we can define a first home space $HS$ that concisely describe how tokens are distributed over places. 
From this token distribution, we can analyze what transition are enabled in order to prove that a specific given marking $q$ ($q_0$ being the usual case) is always reachable from any element of $HS$. 
When this is possible, it can easily be deduced that $q$ is a home state. 
Then, it may be possible using property \ref{prop: home-state-liveness} to prove which transition are live and whether the Petri Net is live or not. 
This will be illustrated later with a few examples in section \ref{sec: ex}.

\section{Reasoning with invariants, semiflows, and home spaces}
\label{sec: ex}

Invariants, semiflows, and home spaces can be used to prove a rich array of behavioral properties of conceptual models such as labeled transition system or Petri Net even within different settings, in particular when using parameters. 

We used to prove liveness by starting by a known home space then proceeding cases by cases, sub-case by sub-cases using a generating set of semiflows, we can often prove that the initial state is a home state (definition \ref{def: home state}) and from there conclude to the liveness of the model using properties \ref{prop: home-state-liveness} (see example in \cite{VautherinM84,M23}.
Here, through two parameterized examples, we proceed by using basic arithmetic for the first example, or by refinement of home spaces in the second example. This methodology brings us closer to an automated analysis to address liveness in parameterized models.

Analysis can be performed with incomplete information on the initial marking as shown in the first example below or on a subsystem exhibiting some compositionality ability. 
It can be described with parameters which will make the invariant calculus more complex but still tractable as shown in the subsequent examples. 
Most of the time, especially with actual system models, it will be possible to conclude avoiding a painstaking symbolic model checking or a parameterized and complex development of a reachability graph \cite{DRvB01}. At the very least it could be envisioned to combine this method with existing ones in order to get an acceleration. 

\subsection{A tiny example}

\begin{figure}[ht]
\centering
\includegraphics[width=0.6\textwidth]{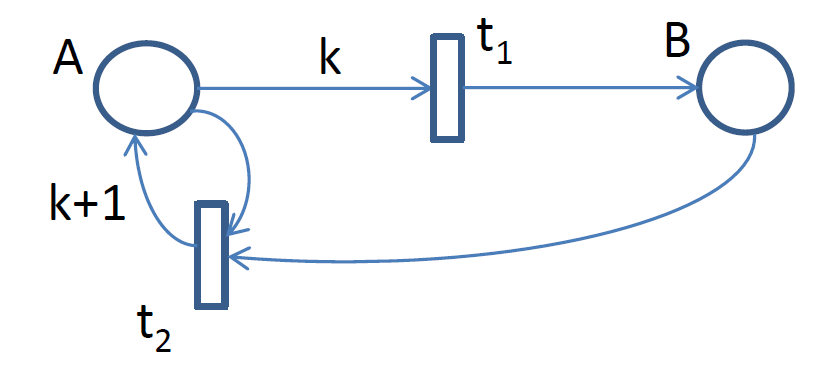}
\caption{This parameterized version TN(k) of our tiny Petri Net is live if and only if $g^\top q_0 > k$ and is not a multiple of $k$, whatever is the initial marking of $B$. For $k=1$, the Petri Net has no live transition whatever is the initial marking. 
}
\label{tinyk}
\end{figure}

The Petri Net $TN = \langle \{A,B\}, \{t_1,t_2\}, Pre, Post \rangle$ in Figure~\ref{tinyk} is defined by: 

$Pre(\cdot,t_1)^\top =(k,0); Pre(\cdot,t_2)^\top =(1,1)$;

$Post(\cdot,t_1)^\top =(0,1); Post(\cdot,t_2)^\top =(k+1,0).$

This example can be first found in \cite{BR82} or in \cite{M83} without proof. Here, the analysis is generalized by introducing a parameter $k$ such that $k>1$.




What is remarkable about the analysis of this tiny example is that it was not necessary to develop a symbolic reachability graph in order to decide whether or not the Petri Net is live or bounded. We could analyze the Petri Net even partially ignoring the initial marking (i.e. considering $q_0(A)$ as an additional parameter and without even considering the value taken by $q_0(B)$ ).



We have the following minimal semiflow: $g^\top  = (1, k)$ and we can show that the TN(k) is not live if and only if $g^\top q_0 \leq k$ or $g^\top q_0 = nk$ where $n \in \mathbb{N}$, independently of $q_0(B)$ which could be considered as a second parameter. 

First, if $g^\top q_0 < k$ then the enabling threshold of $t_1$ can never be reached (property \ref{prop: enabling-threshold}) and neither $t_1$ nor $t_2$ can be executed (since $q_0(B)$ is necessarily null to satisfy the inequality). 
Second, if $g^\top q_0 \geq k$ then we consider the Euclidean division of $g^\top q_0$ by $k$ giving: $g^\top q_0 = nk + i$ where $i < k$ then since $g$ is a semiflow, $g^\top q = q(A)+kq(B) \equiv i\ (mod \ k)$ therefore $q(A) \equiv i\ \forall q \in RS(TN(k),q_0)$. If $i=0$  then we have $q(A)= nk - kq(B)$ and $t_1$ can be enabled $n - q(B)$ times to reach a marking with zero token in $A$. 
If $i \neq 0$ and $g^\top q_0 > k$ then $q(A) \neq 0$ and either $q(A) > k$ or $q(B) \neq 0 \ \forall q \in RS(TN(k),q_0)$. In the first case, $t_1$ is enabled; in the second case $t_2$ is enabled. 
It is easy to conclude that the Petri Net $TN(k)$ is live if and only if $g^\top q_0 > k$ and is not a multiple of $k$ whatever is the initial marking of $B$.
\hfill 
$\square$ 


Even if adding more resources to the system under study can bring confidence that some actions will eventually be performed, this  does not necessarily result in solving a deadlock issue; this could even create a deadlock situation! In this regard, adding more resources may ends up as a ``false good idea” that is encountered with many students. This is certainly one of the reasons why the coverability tree does not allow to study liveness \cite{F93}.

\subsection{A parameterized example from the telecommunication industry}
\label{subsec: example-summary}
\begin{figure}[ht]
\centering
\includegraphics[width=0.9\textwidth]{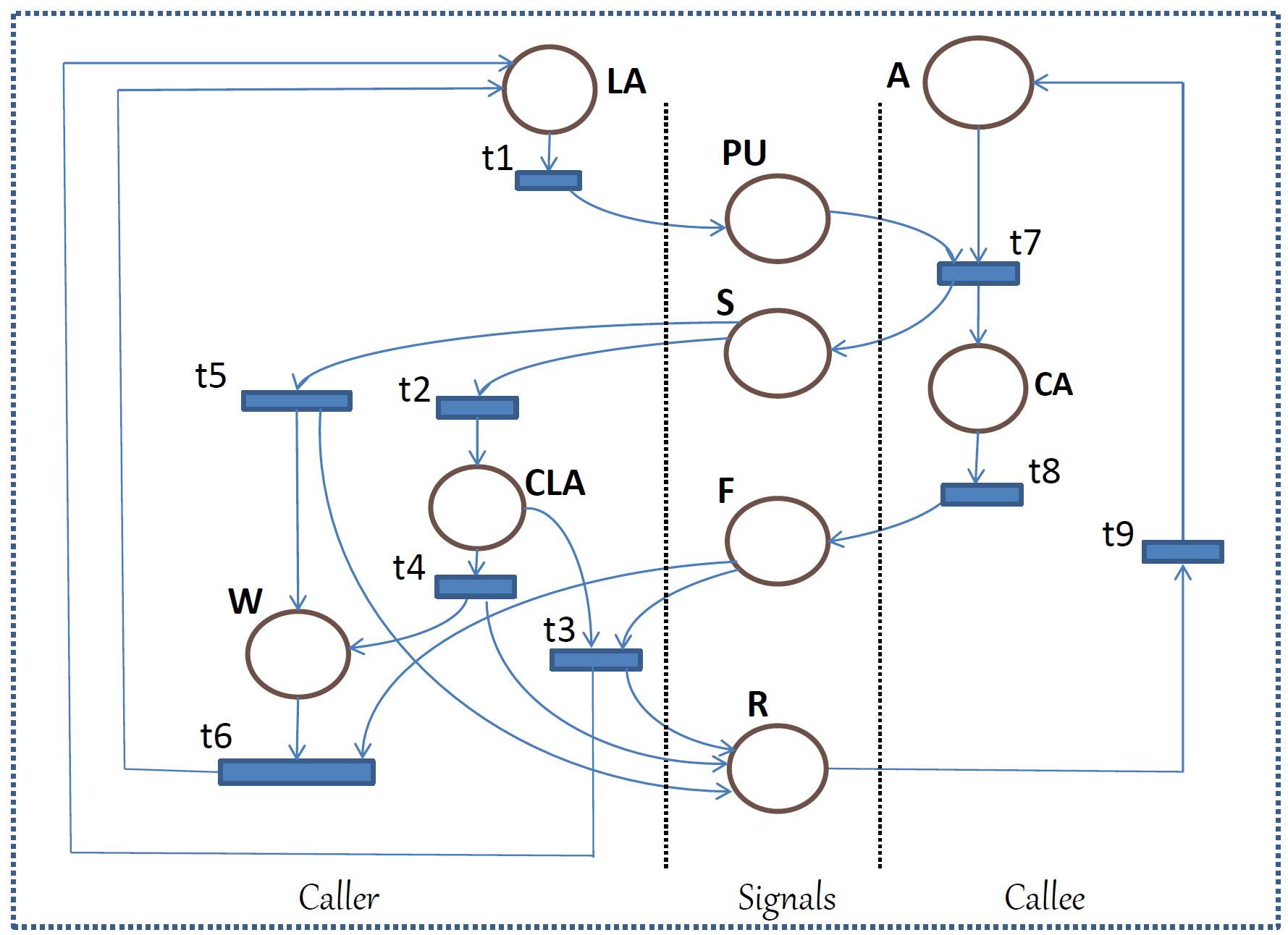}
\caption{This Petri Net TEL(x,y) has exactly three minimal semiflows of minimal support, constituting a generating set over $\mathbb{Q}^+$: $\mathcal{GB}_1 = \{f_1,\ f_2,\ f_3\}$ such that: 
\newline
$f_1(LA)=f_1(CLA)=f_1(WLA)=f_1(PU)=f_1(S)=\ 1$ and $f_1(p)=\ 0$ for any other place,
\newline
$f_2(LA)=f_2(PU)=f_2(F)=f_2(CA)=\ 1$ and $f_2(p)=\ 0$ for any other place,
\newline
$f_3(CLA)=f_3(S)=f_3(R)=f_3(A)= \ 1$ and  $f_3(p)=\ 0$ for any other place. \newline
$\mathcal{GB}_1$ and its corresponding invariants are concisely represented by a tableau Figure \ref{fig: tableau-parameterrized}.
}
\label{Mame}
\end{figure}

The Petri Net $TEL(x,y)= \langle P_{TEL},T_{TEL},Pre,Post \rangle$ is described Figure \ref{Mame}

where $P_{TEL} = \{LA, CLA, WLA, A, CA, PU, S, F, R\}$ 

and $T_{TEL}= \{t_1, t_2, t_3, t_4, t_5, t_6, t_7, t_8, t_9\}$.

TEL(x,y) is a reduced  Petri Net version of a Fifo Net published in \cite{MemFink85,MeMa81}. 
It is well known that reduction rules preserve liveness, however, it is also well known that some rules do not preserve boundedness \footnote{See \cite{BR82,Colom2003} for Petri Nets reduction rules.}.
A sketchy and incomplete version of the analysis hereunder is published in \cite{M23}.

This example is representing two subscriber types, ``a caller" and a ``callee," having a conversation (places $CLA$ and $CA$ respectively).
We simplify the model by taking into account only the actions related to calling for the caller and the actions related to being called for the callee.
Initially, they are in an idle state with places $LA$ and $A$ marked with one token per subscriber. We will want this principle be satisfied for any evolution of TEL(x,y).
Signals $PU$ and $R$ are sent from the caller to the callee and signals  $S$ and $F$ from the callee to the caller. 
The overall desired behavior is that caller and callee cannot go back to their idle state as long they have not received all the signals sent to them despite the fact that they both can hang up at any time making the order in which signals $F$ and $R$ are sent and received undetermined.

From their idle state (place $LA$), the caller can pick up their phone (transitions $t_1$) sending the signal $PU$ to the callee. 
From their idle state (place $A$), the callee, upon receiving the signal $PU$, can pick up their phone (transition $t_7$), send the signal $S$ and go
to conversation (place $CA$) from where they can hang up (transition $t_8$) at any time sending the signal $F$ to the caller. 
Receiving the signal $S$, the caller can go (transition $t_2$) to the conversation (place $CLA$).
They can also hang up at any time (transitions $ t_3
, t_4, t_5$) sending the signal $R$ to the callee.
After hanging up via $t_4$ or $t_5$, the caller will have to wait (place $W$) until they receive the signal $F$ from the callee before going back (transition $t_6$) to their initial idle state $LA$.
The callee can go back to their idle state $A$ only upon receiving the signal $R$ (transition $t_9$).

The initial state $q_0$ is such that $q_0(LA)= x, q_0(A)=y$ and $q_0(p)=0$ for any other place: the modeled system is in its idle state.

\subsubsection{Proving the TEL(x,y) example with parameters}
\begin{figure}[ht]
\centering
\includegraphics[width=0.8\textwidth]{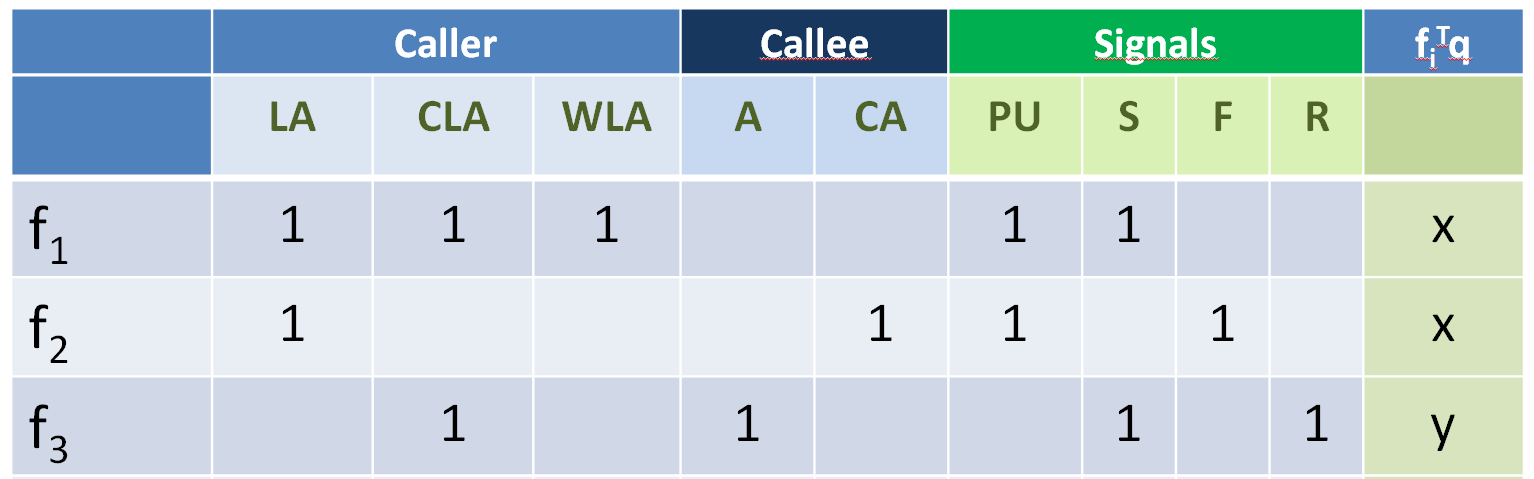}
\caption{This tableau is concise way to represent a generating set and corresponding invariants of $\mathcal{GB}_1$. It reads for instance: $f_1(LA) = 1$ or $f_1(A)=0$. The three semiflows $f_1,\ f_2,\ f_3$ do not depends on the initial state (even enriched with parameters); only the evaluation functions of the three corresponding invariants are depending on the initial state, therefore are defined with parameters.}
\label{fig: tableau-parameterrized}
\end{figure}

A similar analysis scheme could be conducted with a colored Petri Net instead of the Petri Net of Figure \ref{Mame} with two parameters to model $x$ callers and $y$ callees, where $x>0$ and $y>0$. Considering the parameterized Petri Net of Figure \ref{Mame}, 
the initial state $q_0$ becomes: 
$q_0(LA)=x, \ q_0(A)=y$ and $q_0(p)=0$ for any other place $p$. 
The generating set $\mathcal{GB}_1 = \{f_1,\ f_2,\ f_3\}$ is minimal and its three associated invariants are:

$ \forall q \in RS(TEL(x,y), q_0)$:

$f_1^\top q=f_1^\top q_0=x$, 
$f_2^\top q=f_2^\top q_0=x$, 
$f_3^\top q=f_3^\top q_0=y$,

as described Figure \ref{fig: tableau-parameterrized}.

This time, we want to prove that TEL(x,y) is live and satisfies the following property:

$\mathcal{P}_1(x,y) : HS(z) = \{q|\ q(CLA)=q(CA)=z \ \text{and}\ z \leq min(x,y)\} $ is a home space,

$\mathcal{P}_2(x,y) : \forall q \in RS(TEL(x,y), q_0)$,
$q(CLA) \leq min(x,y)$ and 
$q(CA) \leq min(x,y)$.

First, from property \ref{prop: inter-home-spaces} that $HS_0 = \{q \in Q\ |\ f_1^\top q=f_2^\top q=x, \text{and} 
f_3^\top q=y$ \} is a home space.

From any state $q \in HS_0$, it is always possible to execute the sequence:

$\sigma_1=(t_5t_9)^{q(S)}(t_4t_9)^{q(CLA)}t_9^{q(R)}$.
We then reach a state $q_1$ such that $q_1(S)=q_1(CLA)=q_1(R)=0\ $ defining a second home space $HS_1= \{q \in HS_0 \ |\ q(S)=q(CLA)=q(R)=0\}$.
From the invariant associated with $f_3$, it can be directly deduced that $\forall q_1 \in HS_1$, we have $q_1(A)=y$.
Similarly, it is always possible to empty $CA$ from its tokens: $\forall q_1 \in HS_1$, it is always possible to execute
$\sigma_2=t_8^{q(CA)}$ reaching a state $q_2$ such that $q_2(CA)=0$ defining a third home space $HS_2= \{q \in HS_1 \ |\ q(CA)=0\}$.

Since $y>0$, from any state $q_2$ in $HS_2$, it is always possible to execute the sequence: 

$\sigma_3=(t_7t_5t_9t_8)^{q_2(PU)}$.
We then reach a state $q_3$ still in $HS_2$ (each time $t_7$ puts a token in $S$ and $CA$, we can execute $t_5t_9$ and $t_8$ respectively to return to $HS_2$) such that $q_3(PU)=0$ defining a fourth home space $HS_3= \{q \in HS_2 \ |\ q(PU)=0\}$.
From the two invariants associated with $f_1$ and $f_2$, it can be deduced that $\forall q \in HS_3$, we have $q(LA)+q(WLA)=x$ and $q(LA)+q(F)=x$ respectively, hence necessarily, $q(WLA)=q(F)$. 
Therefore, from any state $q_3$ in $HS_3$, it is always possible to execute the sequence: $\sigma_4=t_6^{q_3(F)}$ reaching a state $q_4$ still in $HS_3$ such that $q_4(WLA)=q_4(F)=0$ therefore, $q_4(LA)=x$. 
The only such state is $q_0$ which being always reachable via the sequence $\sigma_1\sigma_2\sigma_3\sigma_4$, is a home state.
From there, $t_1$ is live (from property \ref{prop: home-state-liveness}) and it becomes easy to prove that the Petri Net is live.

\begin{figure}[ht]
\centering
\includegraphics[width=0.8\textwidth]{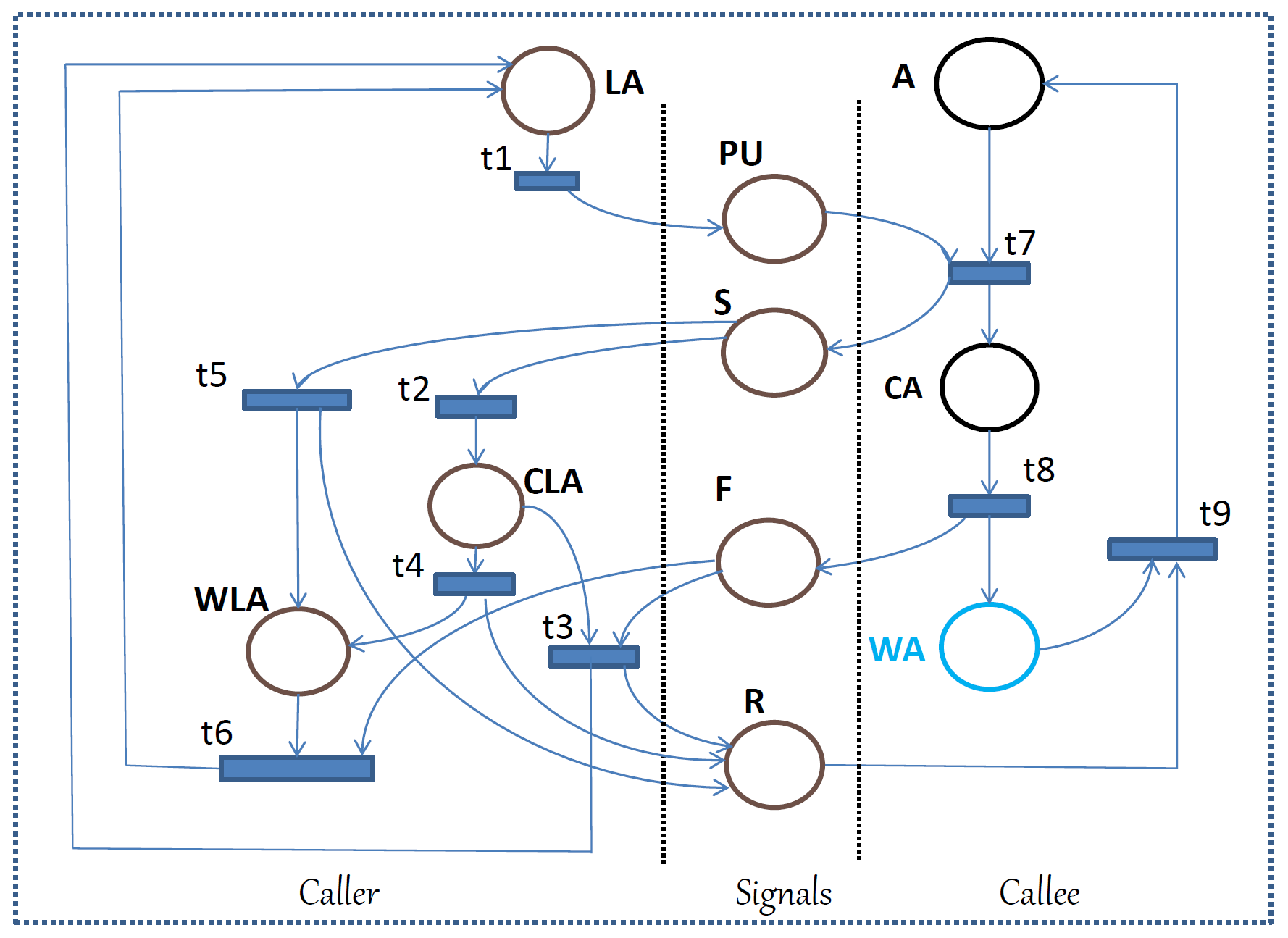}
\caption{TEL2(x,y): By adding the place $WA$ and connecting it to the transitions $t_8$ and $t_9$, we generate a fourth minimal semiflow without changing the three first ones.
\newline
$f_4(A)=f_4(CA)=f_4(WA)=\ 1$ and $f_1(p)=\ 0$ for any other place}
\label{fig: mame2}
\end{figure}

From $q_0$ home state, we can always execute the sequence $\sigma_z = (t_1t_7t_2)^z$ as long as $z \leq min(x,y)$; therefore
$HS(z) = \{q|\ q(CLA)=q(CA)=z \ \text{and}\ z \leq min(x,y)\} $ is a home space meaning that it is always possible to have $z$ pairs of subscribers in a conversation at the same time which satisfies $\mathcal{P}_1(x,y)$.

Furthermore, by applying theorem \ref{th: bounds} using $\mathcal{GB}_1$, we directly have:

$\forall q \in RS(TEL(x,y),q_0), q(CLA) \leq \mu(CLA,q_0)= min(x,y)$.

This is the first half of property $\mathcal{P}_2(x,y)$ and it means that we can only have up to $min(x,y)$ simultaneous conversations.

However, the same theorem allows deducing not only that:

$\forall q \in RS(TEL(x,y),q_0), q(CA) \leq \mu(CA,q_0)= x$,
but also that there is not hope to find a better bound for $CA$ with the only help of semiflows.
Worse, it can be shown that $\mu(CA,q_0)= x$ can be reached and that $\mathcal{P}_2(x,y)$ is not satisfied. In the case where $x > y$, this would model the fact that a callee can be in a simultaneous conversation with several callers (which is a possibility today, but was not expected here). Moreover, this situation does not comply with the principle of one token per subscriber since we would have more than $y$ callee.

This situation is due to an oversimplification of TEL(x,y). By restoring one reduction rule back and remembering that in particular, the callee process was modeled by state machines, we construct the Petri Net $TEL2(x,y)$ of Figure \ref{fig: mame2}.
$TEL2(x,y)$ has an augmented generating set $\mathcal{GB}_2 = \mathcal{GB}_1 \cup \{f_4$\} where $f_4$ is a new minimal semiflow such that $f_4(A)=f_4(CA)=f_4(WA)=\ 1$ and $f_1(p)=\ 0$ for any other place and is directly associated with the following invariant: $\forall q \in RS(TEL2(x,y),q_0),\ f_4^\top q=f_4^\top q_0=y$.  
We can apply theorem \ref{th: bounds} again to $CA$ and this time obtain 
$\forall q \in RS(TEL2(x,y),q_0),\ q(CA) \leq \mu(CA,q_0) = min(x,y)$ and finish to prove $\mathcal{P}_2(x,y)$. Indeed, it can be shown that with similar method of proofs, $TEL2(x,y)$ satisfies $\mathcal{P}_1(x,y)$ and is live.
\hfill 
$\square$

\section{Conclusion}
\label{sec: concl}
As soon as we can associate the behavior of a system under study with a set of state variables and follow its evolution through sequences of states forming a reachability graph (or a structure) where each edge is associated with a transition, we can assume the existence of an underlying Transition System. Doing so, invariants and home spaces can be checked or proven along all possible sequences and help understanding and verifying the functioning of such a system.

It has been recalled how semiflows create a link from the static topology (the bipartite graph) to the dynamic evolution (the variation of the number of tokens) of the Petri Net. 
They support constraints over all possible markings which greatly help analyzing and discovering behavioral properties (even some unspecified ones). 
Semiflows infer a class of invariants with a constant evaluation function that can be computed from the initial marking of the Petri Net under consideration. 
Last but not least, they can be characterized by a generating set that can be used in order to support some elegant level of behavioral analysis even with the presence of some level of parameterization.






We believe that these results may  be enriched along two different alleys. 
From a mathematical point of view, the relation with integer linear programming or convex geometry has been investigated many times, particularly in \cite{ColomS89}, however, we believe it could be fruitful to look at the notion of toric varieties and saturated semigroups \cite{Oda12}. 
From a Petri Net and, even more broadly, from a transition system theory point of view, applying these new results to a variety of models (for example, colored Petri Nets or any transition system that can be associated with a system of equations such as Equation (\ref{eq:inv-semiflow})) remains to be done. 

Finally, we would like to stress the possibility to automate the proof scheme exhibited throughout the parameterized examples of Section \ref{sec: ex} in particular by introducing tableaux (facilitating calculation) and a method by refinements of home space which constitute two new steps towards this goal bringing simplification and clarification compared to proceeding by cases and sub-cases (used in \cite{M23} and other papers). 

\bibliography{bibfile}

\newcommand{\etalchar}[1]{$^{#1}$}
\begin{thebibliography}{CMPAW09}

\bibitem[AB86]{AB86}
N.~Alon and K.~A. Berman.
\newblock Regular hypergraphs, gordon’s lemma, steinitz’ lemma and invariant theory.
\newblock {\em \textit{J. of Combinatorial Theory, A}}, 43:pp. 91--97, 1986.

\bibitem[BEI{\etalchar{+}}20]{BEISW20}
M.~Bozga, J.~Esparza, R.~Iosif, J.~Sifakis, and C.~Welzel.
\newblock Structural invariants for the verification of systems with parameterized architectures.
\newblock In A.~Biere and D.~Parker, editors, {\em \textit{Tools and Algorithms for the Construction and Analysis of Systems}}, pages 228--246, Cham, (2020). Springer Int. Pub.

\bibitem[Bra82]{BR82}
G.~W. Brams.
\newblock {\em R{\'e}seaux de Petri: Th{\'e}orie et Pratique}.
\newblock Masson, Paris, France, 1982.

\bibitem[CMPAW09]{CMPW09}
G.~Ciardo, G.~Mecham, E.~Paviot-Adet, and M.~Wan.
\newblock P-semiflow computation with decision diagrams.
\newblock In Wolf K.~(eds) Franceschinis~G., editor, {\em ATPN, Petri Nets 2009}, LNCS 5606, pages 143--162. Springer, Berlin, Heidelberg, (2009).

\bibitem[CS89]{ColomS89}
J.~M. Colom and M.~Silva.
\newblock Convex geometry and semiflows in {P/T} nets. {A} comparative study of algorithms for computation of minimal p-semiflows.
\newblock In G.~Rozenberg, editor, {\em \textit{Advances in Petri Nets 1990 [$10^{th}$ Int. Conf. on ATPN}, Bonn, Germany, Proc.]}, LNCS 483, pages 79--112. Springer, 1989.

\bibitem[CST03]{ColomTS2003}
J.~M. Colom, M.~Silva, and E.~Teruel.
\newblock Properties.
\newblock In C.~Girault and R.~Valk, editors, {\em \textit{Petri Nets for Systems Engineering: A Guide to Modeling, Verification, and Applications}}, pages 53--72. Springer Berlin Heidelberg, 2003.

\bibitem[CTSH03]{Colom2003}
J.~M. Colom, E.~Teruel, M.~Silva, and S.~Haddad.
\newblock Structural methods.
\newblock In C.~Girault and R.~Valk, editors, {\em \textit{Petri Nets for Systems Engineering, A guide to Modeling, Verification, and Applications}}, pages 277--316. Springer Berlin Heidelberg, 2003.

\bibitem[DL16]{DworLo16}
L.~W. Dworzanski and I.~A. Lomazova.
\newblock Structural place invariants for analyzing the behavioral properties of nested petri nets.
\newblock In F.~Kordon and D.~Moldt, editors, {\em \textit{ATPN and Concurrency}}, pages 325--344, Cham, 2016. Springer Int. Pub.

\bibitem[DRvB01]{DRvB01}
G.~Delzanno, J.~F. Raskin, and L.~van Begin.
\newblock Attacking symbolic state explosion.
\newblock In {\em \textit{Proc. of the $13^{th}$ International Conference on Computer Aided Verification, CAV 2001}}, LNCS 2102, pages 298--310. Springer, (2001).

\bibitem[FEJ89]{EJ89}
D.~Frutos~Escrig and C.~Johnen.
\newblock Decidability of home space property.
\newblock Technical report, Université Paris-Saclay, LRI, UA CNRS 410, RR n 503, 1989.

\bibitem[FH24]{FinHil24}
A.~Finkel and M.~Hilaire.
\newblock Resilience and home-space for wsts.
\newblock In Rayna Dimitrova, Ori Lahav, and Sebastian Wolff, editors, {\em Verification, Model Checking, and Abstract Interpretation}, pages 147--168, Cham, 2024. Springer Nature Switzerland.

\bibitem[Fin93]{F93}
A.~Finkel.
\newblock The minimal coverability graph for petri nets.
\newblock {\em in W. Reisig and G. Rozenberg (Eds) \textit{Advances in Petri Nets}, LNCS 674}, pages pp. 210--243, 1993.

\bibitem[GV03]{GV03}
C.~Girault and R.~Valk.
\newblock {\em Petri Nets for Systems Engineering, A guide to Modeling, Verification, and Applications}.
\newblock Springer, 2003.

\bibitem[HDK14]{HDMK14}
Thomas Hujsa, Jean{-}Marc Delosme, and Alix~Munier Kordon.
\newblock Polynomial sufficient conditions of well-behavedness and home markings in subclasses of weighted petri nets.
\newblock {\em {ACM} Trans. Embed. Comput. Syst.}, 13(4s):141:1--141:25, 2014.

\bibitem[JACB18]{JACB18}
M.~D. Johnston, D.~F. Anderson, G.~Craciun, and R.~Brijder.
\newblock Conditions for extinction events in chemical reaction networks with discrete state spaces.
\newblock {\em \textit{J. Mathematical Biology}}, 76,6:pp. 1535–--1558, 2018.

\bibitem[JL22]{JALE22}
P.~Jančar and J.~Leroux.
\newblock Semilinear home-space is decidable for petri nets.
\newblock {\em arXiv 2207.02697}, 2022.

\bibitem[KJ87]{Kruck86}
F.~Kr{\"u}ckeberg and M.~Jaxy.
\newblock Mathematical methods for calculating invariants in petri nets.
\newblock In G.~Rozenberg, editor, {\em \textit{Advances in Petri Nets 1987}}, pages 104--131. Springer Berlin Heidelberg, 1987.

\bibitem[Lan02]{Lan02}
S.~Lang.
\newblock {\em Algebra, $3^{rd}$ revised Ed.}
\newblock GTM, Springer, 2002.

\bibitem[Mem77]{M77}
G.~Memmi.
\newblock Semiflows and invariants. application in petri nets theory.
\newblock {\em in \textit{Journ{\'e}es d'Etudes sur les R{\'e}seaux de Petri AFCET-Institut de Programmation)}}, pages pp. 145--150, 1977.

\bibitem[Mem78]{M78}
G.~Memmi.
\newblock {\em Fuites et Semi-flots dans les R{\'e}seaux de Petri}.
\newblock Th{\`e}se de Docteur-Ing{\'e}nieur, U. P. et M. Curie, Paris, France, 1978.

\bibitem[Mem83]{M83}
G.~Memmi.
\newblock {\em Methodes d'analyse de R{\'e}seaux de Petri, R{\'e}seaux a Files, Applications au temps reel}.
\newblock Th{\`e}se d'Etat, U. P. et M. Curie, Paris, France, 1983.

\bibitem[Mem23]{M23}
Gerard Memmi.
\newblock Minimal generating sets for semiflows.
\newblock In Marieke Huisman and Ant{\'o}nio Ravara, editors, {\em Formal Techniques for Distributed Objects, Components, and Systems}, pages 189--205, Cham, 2023. Springer Nature Switzerland.

\bibitem[MF85]{MemFink85}
G.~Memmi and A.~Finkel.
\newblock An introduction to fifo nets-monogeneous nets: {A} subclass of fifo nets.
\newblock {\em \textit{Theor. Comput. Sci.}}, 35:191--214, 1985.

\bibitem[MM81]{MeMa81}
R.~Martin and G.~Memmi.
\newblock Specification and validation of sequential processes communicating by fifo channels.
\newblock In {\em \textit{$4^{th}$ Int. Conf. on Software Engineering for Telecommunication Switching Systems}, Warwick U. Conventry, U.K}, pages 54--57. SIEE, 1981.

\bibitem[Oda12]{Oda12}
T.~Oda.
\newblock {\em Convex Bodies and Algebraic Geometry (An Introduction to the Theory of Toric Varieties)}.
\newblock Springer, 2012.

\bibitem[Sif78]{Si78}
J.~Sifakis.
\newblock Structural properties of petri nets.
\newblock In J.~Winkowski, editor, {\em \textit{MFCS 1978, Proc. $7^{th}$ Symp.}, Zakopane, Poland}, volume~64 of {\em LNCS}, pages 474--483. Springer, (1978).

\bibitem[STC98]{STC1998}
M.~Silva, E.~Teruel, and J.~M. Colom.
\newblock {\em Linear algebraic and linear programming techniques for the analysis of place/transition net systems}, pages 309--373.
\newblock Springer Berlin Heidelberg, 1998.

\bibitem[VJ85]{ValkJ85}
R{\"{u}}diger Valk and Matthias Jantzen.
\newblock The residue of vector sets with applications to decidability problems in petri nets.
\newblock {\em Acta Informatica}, 21:643--674, 1985.

\bibitem[VM84]{VautherinM84}
J.~Vautherin and G.~Memmi.
\newblock Computation of flows for unary-predicates/transition-nets.
\newblock In G.~Rozenberg, H.~Genrich, and G.~Roucairol, editors, {\em \textit{Advances in Petri Nets 1984, European Workshop on Applications and Theory in Petri Nets}, selected papers}, volume 188 of {\em LNCS}, pages 455--467. Springer, 1984.

\bibitem[Wol19]{Wol2019}
K.~Wolf.
\newblock {\em How Petri Net Theory Serves Petri Net Model Checking: A Survey}, pages 36--63.
\newblock Springer\textit{} Berlin Heidelberg, 2019.

\end{thebibliography}

\end{document}